\documentclass[fleqn,10pt]{wlscirep}

\title{Optical Properties of Synthetic Cannabinoids with Negative Indexes }

\author[1,*]{Yao Shen}
\author[2]{Yu-Zhu Chen}

\affil[1]{School of Forensic Science, People's Public Security
University of China, Beijing 100038, PR China}

\affil[2]{Department of Physics, Tianjin University, Tianjin 300350, PR China}

\affil[*]{shenyaophysics@hotmail.com}

\begin{abstract}
Some kinds of psychoactive drugs have the structures which are called
split-ring resonators (SRRs). SRRs might result in negative permittivity
and permeability simultaneously in electromagnetic field. Simultaneous negative indexes
can lead to the famous phenomenon of negative refraction. This optical
property makes it possible to distinguish synthetic cannabinoids
from other abusive psychoactive drugs in the UV-vis region. This optical method
is non-damaged and superior in forensic science. In this paper,
we use tight-binding model calculating the permittivity and permeability
of the main ingredients of synthetic cannabinoids. At the same time,
we give two more results of zolpidem and caffeine. Further we discuss
the negative refraction of the category of zepam qualitatively. \end{abstract}

\begin{document}

\flushbottom
\maketitle

\thispagestyle{empty}


In 1968, Veselago originally introduced the concept of negative index
material (NIM) which exhibit unusual optical property\cite{x1}. For this kind of material, the permittivity
and permeability can be negative simultaneously in electromagnetic
field, which might bring about the phenomenon of negative refraction.
Pendry \textit{et al.} \cite{x2,x3,x4,x5,Luo13} developed Veselago's
theory, their research pointed out that the configuration of split-ring
resonators (SRRs) \cite{x3} which had non-trivial symmetry breaking
\cite{Fang15} is an effective way to realize negative refraction.
Negative refraction has became more and more popular these decades, new
researches and applications come to practice. For instance, the perfect
lenses which are made by NIM can focus all Fourier components of image,
both their propagating and evanescent waves are amplified, the quality
of image and the detection sensitivity are improved\cite{x6-1,x6-2,x6-3,x6-4,Ma10}.
After the theoretical derivation, Shelby \textit{et al.} \cite{x6}
gave the experimental realization two years later. According to Veselago's
theory, molecules with SRRs configurations such as extended metal
atom chains (EMACs), have led a fresh direction of physical and chemical
research \cite{x7,x71,x72,x73,x8,x9,x10,x10-1,x10-2,x10-3,x10-4,x10-5,x10-6,x10-7,x10-10,x10-11,x10-12,x10-13,x10-14,x10-15,me-1,me-2,me-3,Zhang14}.
Recently many fabulous works in different fields come forth, such
as invisible cloaking \cite{e1,e2,e2-1,e2-2,e2-3,e2-4}, liquid crystal
magnetic control \cite{e3}, anapole moment \cite{e4}, detection
of latent fingermarks \cite{newme,add1,add2}, etc. Inspired by these ideas,
negative refraction plays more and more important role in forensic science field since it is a non-damaged
optical method to test materials. More importantly, a small categories
of molecules could show negative respondence to eletromagnetic fields
\cite{Dong15}. Utilizing this property, these categories of molecules
can be distinguished from others. This is significant to differentiate
psychoactive drugs in forensic science. As a matter of fact, according
to the theoretical derivation and calculation, we point out that the
main ingredients of synthetic cannabinoids, the category of zepam,
zolpidem and caffeine show various degree of negative refraction while
other psychoactive drugs can not. We discuss the result of synthetic
cannabinoids, zolpidem and caffeine quantitatively and the category
of zepam qualitatively.

This paper is organized as following: In section~II, we previously
derive the permittivity and permeability of the main ingredients of
synthetic cannabinoids, zolpidem and caffeine theoretically. Subsequently
in section~III, the numerical results of permittivity and permeability
are discussed. Finally, the main results are concluded and future
work is predicted in section~IV.

\section*{Theoretical Derivation}

\label{sec:theory}

\subsection*{H\"{u}ckel Model and Tight-binding Method}

According to Pendry's research, molecules which have the broken ring
configurations can realize special optical property which is called
negative refraction. We find that the main ingredients of synthetic
cannabinoids such as JWH-018 (K2), JWH-073 (K2), AKB48 (K3) and AM2201
(K3) have the architecture of SRRs, therefore, their permittivity
and permeability can be negative simultaneously in electromagnetic
field. Moreover, two more compounds zolpidem and caffeine can also
reveal negative refraction. In this section, we mainly discuss three
kinds of model of synthetic cannabinoids, zolpidem and caffeine.

\begin{figure}
\includegraphics[clip]{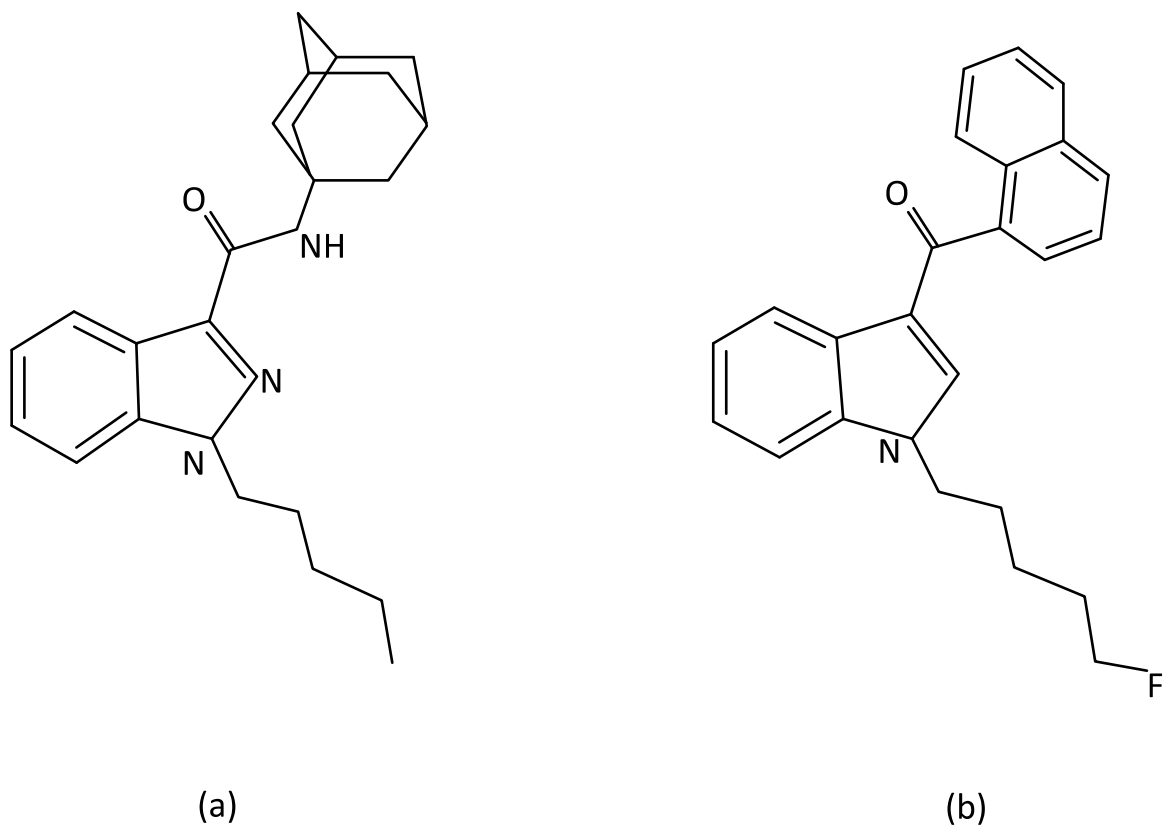}\caption{(a)The molecule of AKB48 (Apinaca, $C_{23}H_{31}N_{3}O$). AKB48 is
N-(1-adamantyl)-1-pentyl-1H-indazole-3-carboxamide. (b)The molecule
of AM2201 ($C_{24}H_{22}FNO$). AM2201 is 1-(5-fluoropentyl)-3-(1-naphthoyl)indole.
\label{fig:two}}
\end{figure}

Figure \ref{fig:two}(a) demonstrates the configuration of AKB48 (apinaca)
molecule, and Figure \ref{fig:two} (b) is the structure called AM2201.
The systematic name of AKB48 is N-(1-adamantyl)-1-pentyl-1H-indazole-3-carboxamide),
and that of AM2201 is 1-(5-fluoropentyl)-3-(1-naphthoyl)indole. They
are the main ingredients of K3 (synthetic cannabinoids).

\begin{figure}
\includegraphics[bb=0bp 0bp 457bp 183bp,clip]{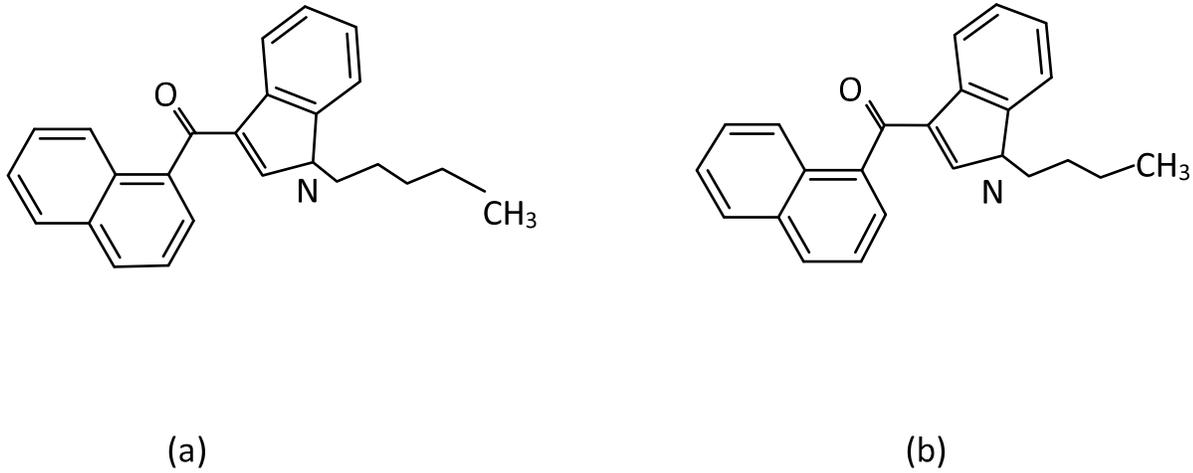}\caption{(a)The molecule of JWH-018. JWH-018 is 1-pentyl-3-(1-naphthoyl)indole
($C_{24}H_{23}NO$). (b)The molecule of JWH-073. JWH-073 is naphthalen-1-yl-(1-butylindol-3-yl)methanone
($C_{23}H_{21}NO$). \label{fig:two-1}}
\end{figure}

Figure \ref{fig:two-1}(a) shows the structure of JWH-018 molecule,
Figure \ref{fig:two-1}(b) gives us JWH-073 molecule. JWH-018 is 1-pentyl-3-(1-naphthoyl)indole
and JWH-073 is naphthalen-1-yl-(1-butylindol-3-yl)methanone. They
are the main ingredients of K2 (synthetic cannabinoids).

Both K2 and K3 are synthetic cannabinoids. The central parts of them
are the broken ring of pentagons. These broken pentagons contribute
to the negative refraction mostly, because all these phenomena come
from the $\pi$ electrons of conjugate ring of these molecules which
are called single-nitrogen-substituted heterocyclic annulenes \cite{x11}.
For AKB48, two carbon atoms of the pentagon are substituted with two
nitrogen atoms, while for AM2201, JWH-018 and JWH-073, only one carbon
atom of the pentagon is replaced by one nitrogen atom. The simplified
model are given in following Figure \ref{three}. The left figure
is the simplified model of AKB48 and the ringt one is the model of
AM2201, JWH-018 and JWH-073. All the atoms of the pentagon are in
the same plane and be marked as Figure \ref{three}. 

\begin{figure}
\includegraphics[clip]{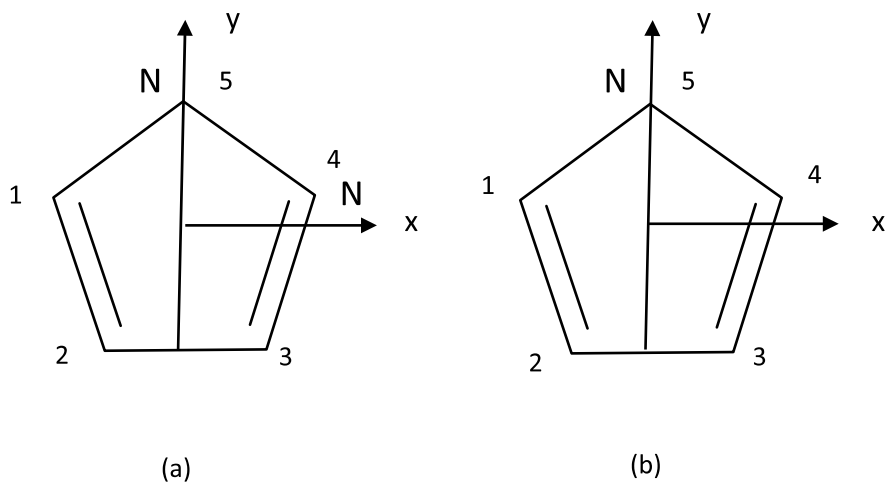}\caption{The Simplified models of synthetic cannabinoids. (a) The model of
AKB48 is 1H-pyrazole ($C_{3}H_{4}N_{2}$) on the left. (b) The models
of AM2201, JWH-018 and JWH-073 are all pyrrole ($C_{4}H_{5}N$) on
the right. The origin is set at the center of the pentagons. \label{three}}
\end{figure}

The simplified model of AKB48 has two nitrogen atoms which are located
at site $4$ and $5$. The other model has only one nitrogen atom
which is placed at site $5$. These two models are called 1H-pyrazole
and pyrrole respectively.

\begin{figure}
\includegraphics[clip]{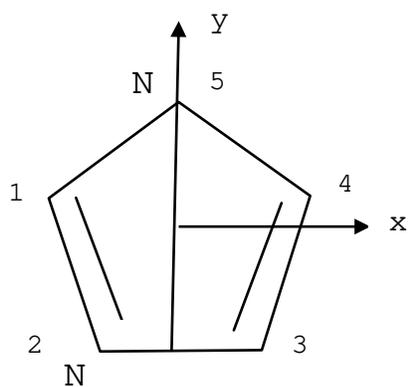}\caption{The Simplified models of zolpidem and caffeine are both 1H-imidazole
($C_{3}H_{4}N_{2}$). 1H-imidazole is different from 1H-pyrazole,
althougt they have the same molecular formula. The nitrogen atoms
are located at different sites. \label{three-1}}
\end{figure}

There are two other cases in psychoactive drugs which also have broken
conjugate ring. The simplified model of zolpidem and caffeine is given
in Figure \ref{three-1}. There are also two nitrogen atoms instead
of two carbon atoms, but their positions are quite different from
1H-pyrazole. Therefore, these dissimilar simplified models show different
degree of negative refraction.

The H\"{u}ckel Hamiltonian of these five $\pi$ electrons conjugate
systems are described as\cite{x12}

\begin{equation}
\mathcal{H}=\sum_{j=1}^{5}\alpha_{j}\left|j\right\rangle \left\langle j\right|+\sum_{j=1}^{5}\beta_{j,j+1}\left(\left|j\right\rangle \left\langle j+1\right|+\left|j+1\right\rangle \left\langle j\right|\right),\label{eq:Hhu}
\end{equation}
where $j$ denotes the site lable, $\alpha$ is site energy and $\beta$
is the resonant integral. When $j=5$, we have $j+1=1$, because of
the ring configuration. We choose $\alpha_{C}=-6.7eV$, $\alpha_{N}=-7.9eV$
\cite{H0}. The resonant integral given by Harrison expression \cite{H1,H2}
\begin{equation}
\beta_{ij}=-0.63\frac{\hbar^{2}}{md_{ij}^{2}}\label{eq:beta}
\end{equation}
where $m$ is the mass of electrons and $d_{ij}$ expresses the bond
length. All data come from NIST \cite{H3}.
\begin{itemize}
\item 1H-pyrazole (AKB48)
\end{itemize}
\begin{equation}
\mathcal{H}=\sum_{j=1}^{3}\alpha_{C}\left|j\right\rangle \left\langle j\right|+\alpha_{N}(\left|4\right\rangle \left\langle 4\right|+\left|5\right\rangle \left\langle 5\right|)+\sum_{j=1}^{5}\beta_{j,j+1}\left(\left|j\right\rangle \left\langle j+1\right|+\left|j+1\right\rangle \left\langle j\right|\right),\label{eq:Hhu-1}
\end{equation}
where $\beta_{1,2}=-2.545eV$, $\beta_{2,3}=-2.393eV$, $\beta_{3,4}=-2.708eV$,
$\beta_{4,5}=-2.632eV$, $\beta_{5,1}=-2.598eV$.
\begin{itemize}
\item pyrrole (AM2201, JWH-018 and JWH-073)
\end{itemize}
\begin{equation}
\mathcal{H}=\sum_{j=1}^{4}\alpha_{C}\left|j\right\rangle \left\langle j\right|+\alpha_{N}\left|5\right\rangle \left\langle 5\right|+\sum_{j=1}^{5}\beta_{j,j+1}\left(\left|j\right\rangle \left\langle j+1\right|+\left|j+1\right\rangle \left\langle j\right|\right),\label{eq:Hhu-2}
\end{equation}
where $\beta_{1,2}=-2.516eV$, $\beta_{2,3}=-2.393eV$, $\beta_{3,4}=-2.516eV$,
$\beta_{4,5}=-2.560eV$, $\beta_{5,1}=-2.560eV$.
\begin{itemize}
\item 1H-imidazole (zolpidem and caffeine)
\end{itemize}
\begin{equation}
\mathcal{H}=\sum_{j=1,3,4}\alpha_{C}\left|j\right\rangle \left\langle j\right|+\alpha_{N}(\left|2\right\rangle \left\langle 2\right|+\left|5\right\rangle \left\langle 5\right|)+\sum_{j=1}^{5}\beta_{j,j+1}\left(\left|j\right\rangle \left\langle j+1\right|+\left|j+1\right\rangle \left\langle j\right|\right),\label{eq:Hhu-3}
\end{equation}
where $\beta_{1,2}=-2.783eV$, $\beta_{2,3}=-2.516eV$, $\beta_{3,4}=-2.582eV$,
$\beta_{4,5}=-2.534eV$, $\beta_{5,1}=-2.582eV$.

\begin{figure}
\includegraphics[clip]{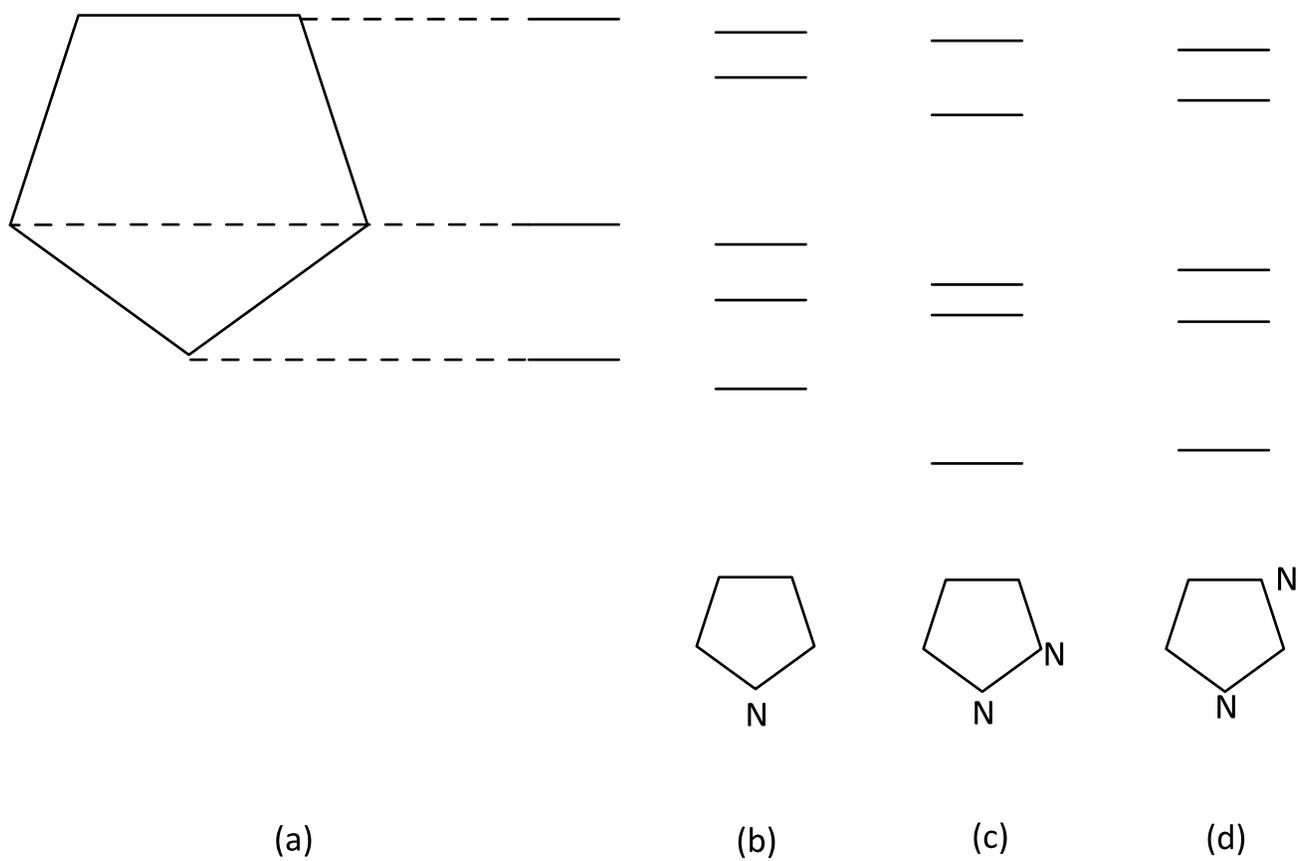}\caption{(a) The energy levels of 1,3-cyclopentadiene ($C_{5}H_{6}$). (b)
The energy levels of pyrrole ($C_{4}H_{5}N$). (c) The energy levels
of 1H-pyrazole ($C_{3}H_{4}N_{2}$). (d) The energy levels of 1H-imidazole
($C_{3}H_{4}N_{2}$). \label{four}}
\end{figure}

By diagonalizing the H\"{u}ckel Hamiltonian~, we calculate the energy
levels of these models as Figure \ref{four}. 

\begin{eqnarray}
\mathcal{H} & = & \sum_{k=1}^{5}\varepsilon_{k}\left|\psi_{k}\right\rangle \left\langle \psi_{k}\right|,
\end{eqnarray}
where
\begin{equation}
\left|\psi_{k}\right\rangle =\sum_{j=1}^{5}C_{ki}\left|j\right\rangle 
\end{equation}
$\varepsilon_{k}$ is the eigen value and $\left|\psi_{k}\right\rangle $
is the eign states. Not losing the generality, we shift all energy
levels by $\alpha_{C}$ for simplicity.
\begin{itemize}
\item 1H-pyrazole (AKB48)
\end{itemize}
$\varepsilon_{1}=-5.762eV$, $\varepsilon_{2}=-2.109eV$, $\varepsilon_{3}=-1.966eV$,
$\varepsilon_{4}=3.465eV$, $\varepsilon_{5}=3.972eV$.
\begin{itemize}
\item pyrrole (AM2201, JWH-018 and JWH-073)
\end{itemize}
$\varepsilon_{1}=-5.330eV$, $\varepsilon_{2}=-1.972eV$, $\varepsilon_{3}=-1.590eV$,
$\varepsilon_{4}=3.708eV$, $\varepsilon_{5}=3.983eV$.
\begin{itemize}
\item 1H-imidazole (zolpidem and caffeine)
\end{itemize}
$\varepsilon_{1}=-5.734eV$, $\varepsilon_{2}=-2.472eV$, $\varepsilon_{3}=-1.743eV$,
$\varepsilon_{4}=3.694eV$, $\varepsilon_{5}=3.855eV$.

All energy levels are nondegenerated. We have five non-interacting
$\pi$ electrons fill in five energy levels. On the basis of Pauli
exclusion principle, the energy of ground state is $E_{0}=2\varepsilon_{1}+2\varepsilon_{2}+\varepsilon_{3}$.
And the second-quantization form of the ground state can be expressed
as 

\begin{equation}
\left|\Psi_{0}\right\rangle =a_{1\uparrow}^{\dagger}a_{1\downarrow}^{\dagger}a_{2\uparrow}^{\dagger}a_{2\downarrow}^{\dagger}a_{3\uparrow}^{\dagger}\left|0\right\rangle ,
\end{equation}
where $\left|0\right\rangle $ is vacuum state and $a_{p\sigma}^{\dagger}$
is the creation operator of the orbital $p$ with spin $\sigma$ ($\sigma=\uparrow,\downarrow$).

We only consider the single-excitation with no flip of electronic
spin. All these three kinds of systems have fourteen single-excitation
states.

The single-excitation states of the system which denote the single
electron excitation from $p$ to $q$ can be assumed as 
\begin{equation}
\left|\Psi_{n}\right\rangle =\left|\Psi_{qp}^{\sigma}\right\rangle =a_{q\sigma}^{\dagger}a_{p\sigma}\left|\Psi_{0}\right\rangle ,
\end{equation}
where $p=1,2,3$, $q=3,4,5$ and $\sigma=\uparrow,\downarrow$. The
coresponding energies are 
\begin{equation}
E_{n}=E_{0}+\varepsilon_{q}-\varepsilon_{p}.
\end{equation}

The ground state and single-excitation states span a subspace. In
this subspace, the Hamiltonian reads

\begin{equation}
\mathcal{H}=\sum_{n=0}^{14}E_{n}\left|\Psi_{n}\right\rangle \left\langle \Psi_{n}\right|.
\end{equation}

\subsection*{Perturbation Theory in Electromagnetic Field}

When the molecules are placed in a time-dependent electromagnetic
field, the total Hamiltonian with the dipole approximation can be
written as
\begin{equation}
H=H_{0}-\vec{\mu}\cdot\vec{E}_{0}\cos(\vec{k}\cdot\vec{r}-\omega t)-\vec{m}\cdot\vec{B}_{0}\cos(\vec{k}\cdot\vec{r}-\omega t),
\end{equation}
where $\vec{k}$ is the wave vector, $\vec{\mu}$ is the electric
dipole moment, $\vec{m}$ is the magnetic dipole moments and $H_{0}=\mathcal{H}$
as defined above is the Hamiltonian without electromagnetic field.
Because the spatial scale of electromagnetic wave length is much longer
than that of the molecules, we assume that the spatial part of the
electromagnetic field can be neglected.

\begin{equation}
H\simeq H_{0}-\vec{\mu}\cdot\vec{E}_{0}\cos(\omega t)-\vec{m}\cdot\vec{B}_{0}\cos(\omega t).
\end{equation}

By a unitary transformation, the Hamiltonian can be simplified into
a time independent one in Schro\"{u}dinger picture. In this picture,
the state and operator are $\left|\Psi^{\prime}\right\rangle =U^{\dagger}\left|\Psi\right\rangle ,$
and $A^{\prime}=U^{\dagger}AU$. The Hamiltonian of the system can
be rewritten as
\begin{eqnarray}
H^{'} & = & U^{\dagger}HU-iU^{\dagger}\dot{U}\nonumber \\
 & \simeq & \sum_{n=1}^{14}E_{n}\left|\Psi_{n}\right\rangle \left\langle \Psi_{n}\right|+\left(E_{0}+\omega\right)\left|\Psi_{0}\right\rangle \left\langle \varPsi_{0}\right|+H^{\prime\prime},
\end{eqnarray}
where
\begin{equation}
U=\exp\left(i\omega\left|\Psi_{0}\right\rangle \left\langle \Psi_{0}\right|t\right),
\end{equation}
 
\begin{eqnarray}
H^{\prime\prime} & = & -\frac{1}{2}\sum_{n=1}^{14}\left[\right.\left(\vec{\mu}_{n0}\cdot\vec{E}_{0}\left|\varPsi_{n}\right\rangle \left\langle \varPsi_{0}\right|+\vec{\mu}_{0n}\cdot\vec{E}_{0}\left|\varPsi_{0}\right\rangle \left\langle \varPsi_{n}\right|\right)\nonumber \\
 &  & +\left(\vec{m}_{n0}\cdot\vec{B}_{0}\left|\Psi_{n}\right\rangle \left\langle \Psi_{0}\right|+\vec{m}_{0n}\cdot\vec{B}_{0}\left|\Psi_{0}\right\rangle \left\langle \Psi_{n}\right|\right)\left.\right],\\
\vec{\mu}_{ns} & = & \left\langle \Psi_{n}\right|\vec{\mu}\left|\Psi_{s}\right\rangle ,\\
\vec{m}_{ns} & = & \left\langle \Psi_{n}\right|\vec{m}\left|\varPsi_{s}\right\rangle .
\end{eqnarray}

Moreover, according to the perturbation theory, the molecular ground
state in Schro\"{u}dinger picture becomes
\begin{equation}
\left|\Psi_{0}^{\prime}\right\rangle =U^{\dagger}\left|\Psi_{0}\right\rangle =\left|\Psi_{0}\right\rangle +\sum_{n=1}^{14}\frac{\left\langle \Psi_{n}\right|H^{\prime\prime}\left|\Psi_{0}\right\rangle }{E_{0}+\omega-E_{n}}\left|\Psi_{n}\right\rangle ,
\end{equation}

\subsubsection*{Derivation of Permittivity}

The electric dipole moment in Schro\"{u}dinger picture is 
\begin{equation}
\vec{\mu}^{\prime}=U^{\dagger}\vec{\mu}U=\sum_{n=1}^{14}\left(\vec{\mu}_{n0}e^{i\omega t}\left|\Psi_{n}\right\rangle \left\langle \Psi_{0}\right|+\vec{\mu}_{0n}e^{-i\omega t}\left|\varPsi_{0}\right\rangle \left\langle \Psi_{n}\right|\right).
\end{equation}
And its expectation value in ground state reads
\begin{equation}
\left\langle \Psi_{0}^{'}\right|\vec{\mu}^{\prime}\left|\Psi_{0}^{'}\right\rangle =-\textrm{Re}\sum_{n=1}^{14}\frac{\vec{\mu}_{n0}\cdot\vec{E}_{0}}{E_{0}+\omega-E_{n}}\vec{\mu}_{0n}e^{-i\omega t}.
\end{equation}
So the electric moment for $N$ identical molecules can be expressed
as 
\begin{equation}
\vec{P}=-\sum_{s=1}^{N}\sum_{n=1}^{14}\frac{\left[\vec{\mu}_{n0}(s)\cdot\vec{E}_{0}\right]\vec{\mu}_{0n}(s)}{E_{0}+\omega-E_{n}}.
\end{equation}

According to the theory of the electric displacement field in the
electromagnetic field, we have 
\begin{equation}
\vec{D}=\varepsilon\vec{E}_{0}=\varepsilon_{0}\varepsilon_{r}\vec{E}_{0}=\varepsilon_{0}\vec{E}_{0}+\frac{\vec{P}}{V},
\end{equation}
where $V$ is the volume and $\vec{P}$ is the electric moment. Thus,
the relative permittivity can be written as
\begin{equation}
\varepsilon_{ij}^{r}\equiv\delta_{ij}-\sum_{s=1}^{N}\sum_{n=1}^{14}\frac{\vec{\mu}_{0n}(s)\cdot\hat{e}_{i}\vec{\mu}_{n0}(s)\cdot\hat{e}_{j}}{\varepsilon_{0}V\left(E_{0}+\omega-E_{n}\right)},\textrm{ for }i,j=x,y,z,\label{eq:ep}
\end{equation}
where $\hat{e}_{i}$ is the unit vector of the lab coordinate system.

The origin of the simplified model is set at the center of the pentagons
(see Figure \ref{three}, \ref{three-1}). As we known, the electric
dipole moment reads
\begin{equation}
\vec{\mu}=-\sum_{j=1}^{5}e\vec{r}_{j},
\end{equation}
where$-e$ is the electric charge and $\vec{r}_{j}$ is the location
vector. Therefore, in the eign states of molecules, the electric dipole
moment operators are

\begin{eqnarray}
\vec{\mu}_{mn}=\left\langle \Psi_{m}\right|\vec{\mu}\left|\Psi_{n}\right\rangle  & = & -e\sum_{j=1}^{5}\left\langle \Psi_{m}\right|\vec{r_{j}}\left|\Psi_{n}\right\rangle =-e\sum_{j=1}^{5}\left\langle \psi_{f}\right|\vec{r_{j}}\left|\psi_{g}\right\rangle =-e\sum_{j=1}^{5}C_{fj}^{*}C_{gj}\vec{r}_{j},
\end{eqnarray}
where $\psi_{g}$ denotes single electron orbital.

\subsubsection*{Derivation of Permeability}

As we known, the magnetic dipole moment can be written as
\begin{equation}
\vec{m}=\frac{-e}{2m_{e}}\vec{L},
\end{equation}
where $\vec{L}$ is the angular momentum of the system. According
to the the Heisenberg equations, we have $L_{x}=L_{y}=0,$
\begin{eqnarray}
\nonumber \\
\nonumber \\
L_{z} & = & xp^{y}-yp^{x}\nonumber \\
 & = & \frac{1}{2}\left(xp^{y}+p^{y}x-yp^{x}-p^{x}y\right)\nonumber \\
 & = & im_{e}\left(x\mathcal{\mathcal{H}}y-y\mathrm{\mathit{\mathcal{H}}}x\right).
\end{eqnarray}
In this case, the magnetic dipole moment can be rewritten as

\begin{eqnarray}
\vec{m} & = & \frac{-e}{2m_{e}}L_{z}\hat{e}_{z}\nonumber \\
 & = & \frac{-ie}{2}\left(x\mathcal{\mathcal{H}}y-y\mathrm{\mathit{\mathcal{H}}}x\right)\hat{e}_{z}\nonumber \\
 & = & \frac{-ie}{2}\sum_{p,q}\sum_{n}E_{n}\left(x_{pn}y_{nq}-y_{pn}x_{nq}\right)\left|\Psi_{p}\right\rangle \left\langle \Psi_{q}\right|\hat{e}_{z},
\end{eqnarray}
where $x_{pn}=\left\langle \psi_{p}\right|x\left|\psi_{n}\right\rangle $,
$y_{nq}=\left\langle \psi_{n}\right|y\left|\psi_{q}\right\rangle $.

Similarly in Schro\"{u}dinger picture, the expectation value of magnetic
dipole moment in ground state is 

\begin{equation}
\left\langle \Psi_{0}^{'}\right|\vec{m}^{\prime}\left|\Psi_{0}^{'}\right\rangle =-\textrm{Re}\sum_{n=1}^{14}\frac{\vec{m}_{n0}\cdot\vec{B}_{0}}{E_{0}+\omega-E_{n}}\vec{m}_{0n}e^{-i\omega t}.
\end{equation}
So the magnetic dipole moment for $N$ identical molecules can be
expressed as
\begin{equation}
\vec{M}=-\sum_{s=1}^{N}\sum_{n=1}^{27}\frac{\mu_{0}\vec{m}_{n0}(s)\cdot\vec{B}_{0}}{E_{0}+\omega-E_{n}}\vec{m}_{0n}(s).
\end{equation}

The magnetic induction in a volume $V$ can be expressed as 
\begin{equation}
\vec{B}=\mu\vec{H}=\mu_{0}\mu_{r}\vec{H}=\mu_{0}\vec{H}+\mu_{0}\frac{\vec{M}}{V},
\end{equation}
where $\vec{H}$ is magnetic field intensity.

The relative permeability of system can be expressed as 
\begin{equation}
\mu_{ij}^{r}\equiv\delta_{ij}-\mu_{0}\sum_{s=1}^{N}\sum_{n=1}^{14}\frac{\vec{m}_{0n}(s)\cdot\hat{e}_{i}\vec{m}_{n0}(s)\cdot\hat{e}_{j}}{V\left(E_{0}+\omega-E_{n}\right)},\textrm{ for }i,j=x,y,z.\label{eq:mu}
\end{equation}

\subsubsection*{Analysis of Analytical Results}

For negative refraction, the permittivity and permeability of the
system should be negative simultaneously. On the basis of the analytical
results given above (\ref{eq:ep}) and (\ref{eq:mu}), the second
parts of these two equations should be greater than unity, even far
exceed. In this case, we reckon that the denominators are much smaller
than the numerator apparently which means $E_{0}+\omega-E_{n}\approx0$
and $\omega\approx E_{n}-E_{0}$. $E_{0}$ is the ground state energy
of initial state. When the molecules are placed in electromagnetic
field, the electromagnetic field gives them a driving frequency $\omega$
which made them transit to the final state with energy $E_{n}$. Here
only single electron excitation is taken into account. The negative
refraction, or in another word, simultaneous negative permittivity
and permeability occurs while the driving frequency of electromagnetic
field $\omega$ approximately equal to the difference of the energy
$E_{n}-E_{0}$ .

\section*{Numerical Simulation of Permittivity and Permeability\label{sec:result}}

\begin{figure}
\includegraphics[scale=1]{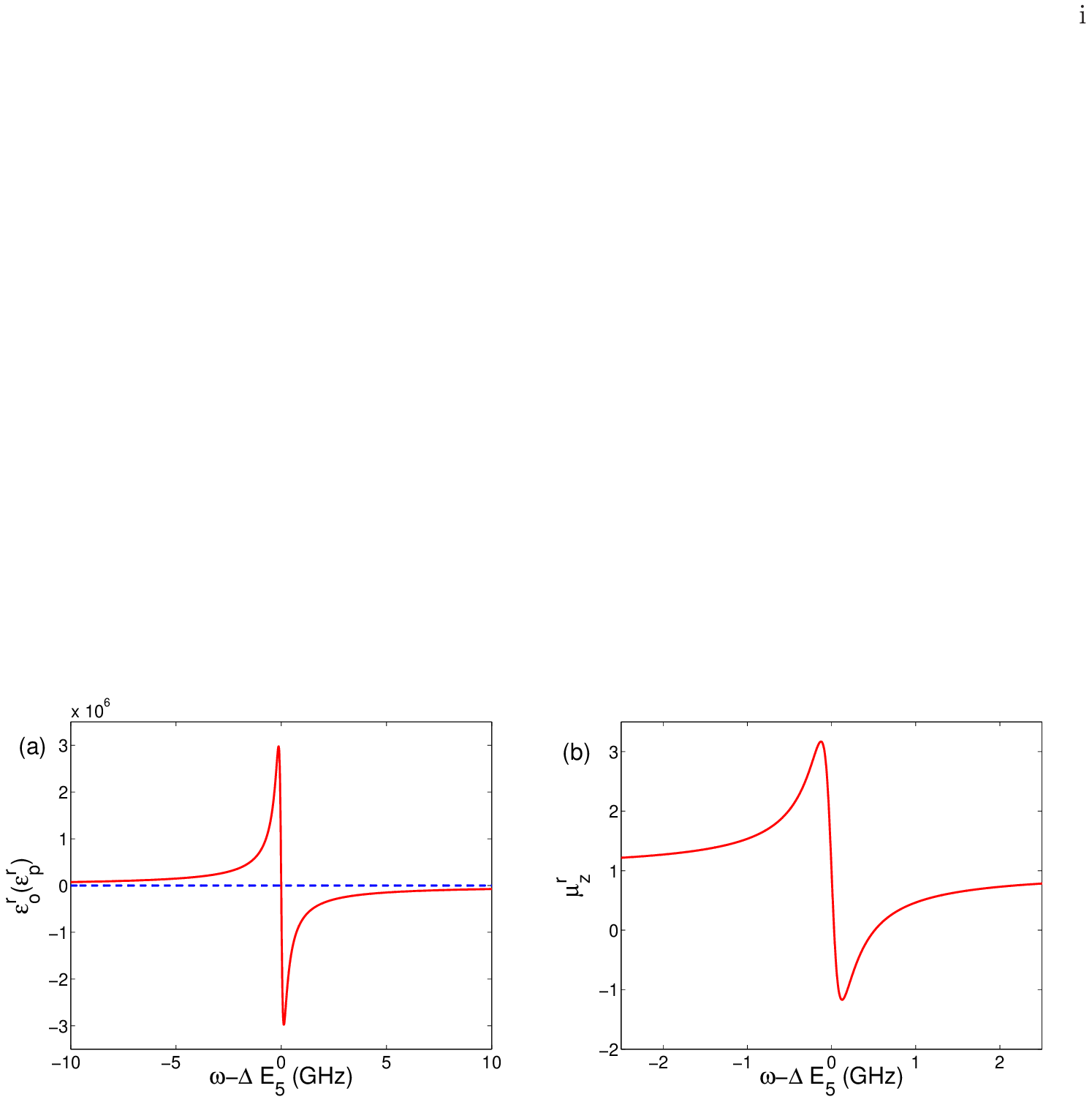}
\caption{The numerical simulation of (a) permittivity and (b) permeability
of pyrrole ($C_{4}H_{5}N$, including AM2201, JWH-018 and JWH-073)
vs the frequency $\omega$ of the electromagnetic field. The axis
$o$ is one main axis in the $xy$ plane while the axis $p$ is another
main axis in this plane. (a) The permittivity $\varepsilon_{o}^{r}$
(red solid line) is negative around the resonance frequency, while
$\varepsilon_{p}^{r}$ (blue dashed line) is always certain constant.
(b) The permeability along $z$ direction $\mu_{z}^{r}$ can be negative
around the resonance frequency simultaneously. \label{fig:re1}}
\end{figure}

\begin{figure}
\includegraphics[scale=1]{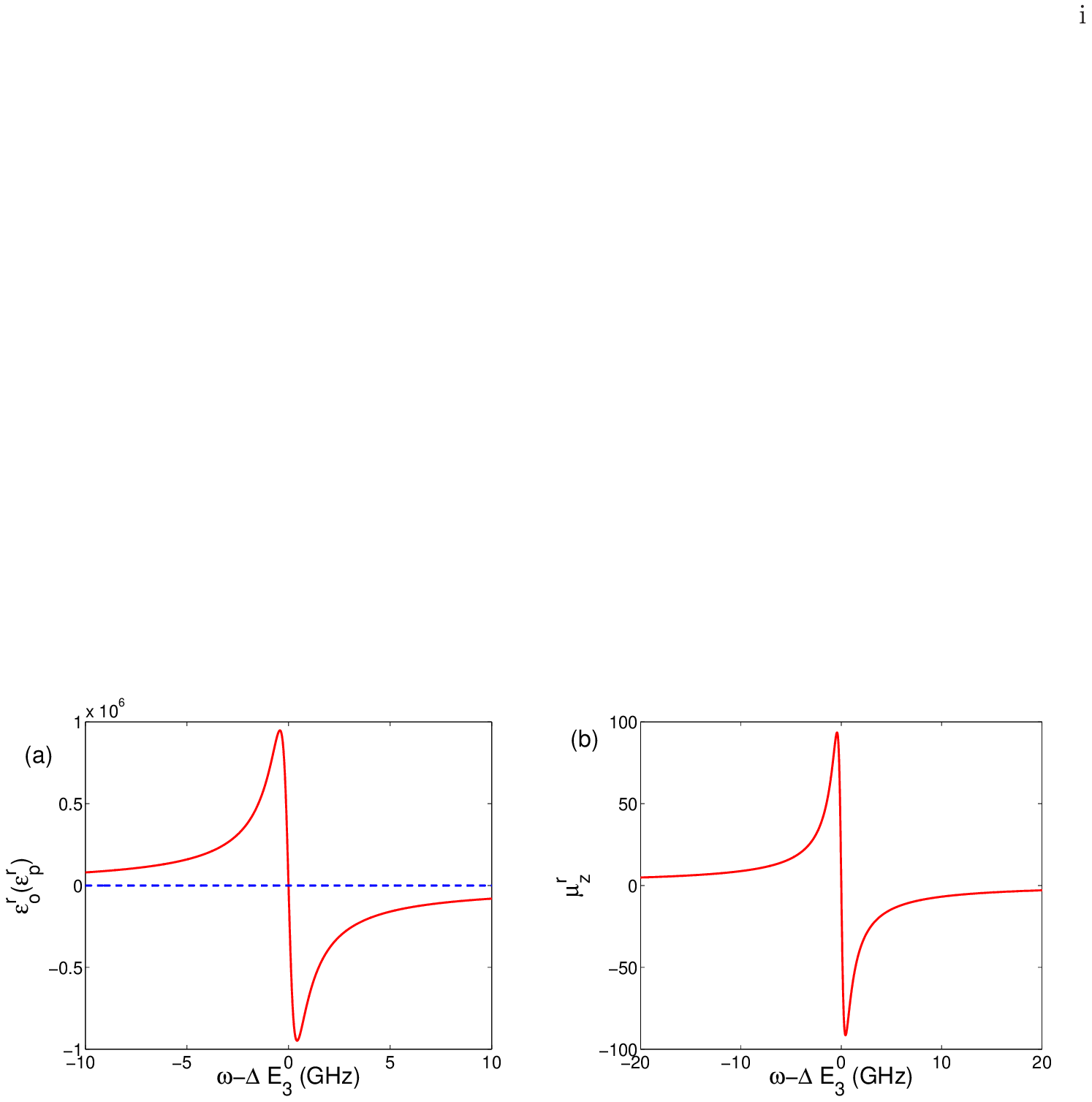}
\caption{The numerical simulation of (a) permittivity and (b) permeability
of AKB48 (1H-pyrazole ($C_{3}H_{4}N_{2}$)) vs the frequency $\omega$
of the electromagnetic field. The axis $o$ is one main axis in the
$xy$ plane while the axis $p$ is another main axis in this plane.
(a) The permittivity $\varepsilon_{o}^{r}$ (red solid line) can be
negative around the resonance frequency. (b) The permeability along
$z$ direction $\mu_{z}^{r}$ can also be negative around the resonance
frequency simultaneously.\label{fig:re2}}
\end{figure}

\begin{figure}
\includegraphics[scale=1]{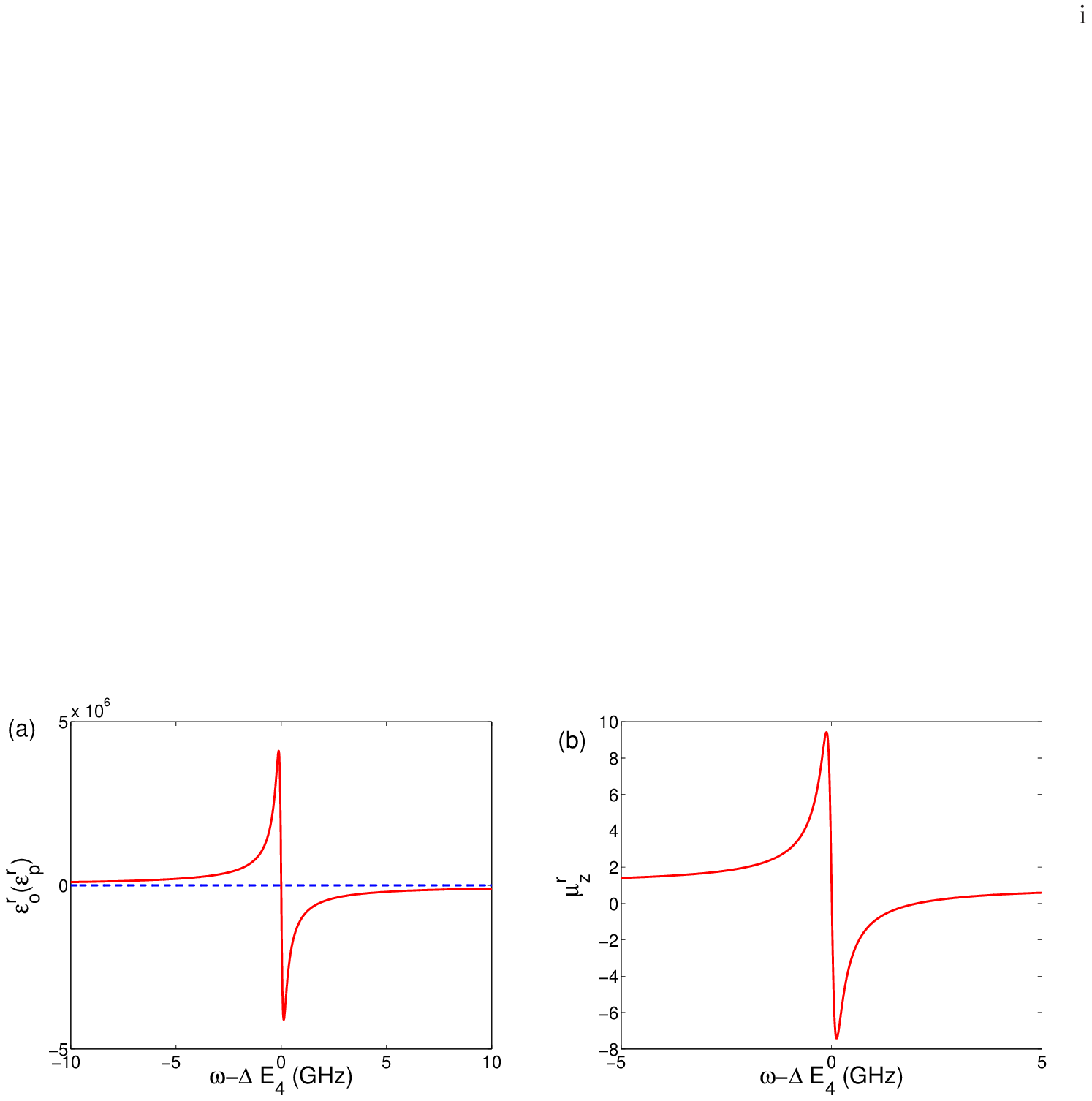}
\caption{The numerical simulation of (a) permittivity and (b) permeability
of zolpidem and caffeine (1H-imidazole ($C_{3}H_{4}N_{2}$)) vs the
frequency $\omega$ of the electromagnetic field. The axis $o$ is
one main axis in the $xy$ plane while the axis $p$ is another main
axis in this plane. (a) The negative permittivity $\varepsilon_{o}^{r}$
(red solid line) and constant $\varepsilon_{p}^{r}$ (blue dashed
line) are shown. (b) The negative permeability along $z$ direction
$\mu_{z}^{r}$ is calso given.\label{fig:re3}}
\end{figure}

The analitical results of section~\ref{sec:theory} suggest that
in certain frequency regime the molecules of certain psychoactive
drugs can demonstrate negative refraction in the UV-vis region. These
psychoactive drugs mainly include synthetic cannabinoids, zolpidem
and caffeine. In this section, we show the numerical results and analyses
to corroborate it. These molecules of psychoactive drugs can be simplified
into three kinds of two dimensional models (see Figure \ref{three}
and Figure \ref{three-1}). Figure \ref{fig:re1} gives the numerical
simulation of permittivity and permeability of pyrrole ($C_{4}H_{5}N$)
which is the simplified model of AM2201, JWH-018 and JWH-073. Figure
\ref{fig:re2} is the result of AKB48 (1H-pyrazole ($C_{3}H_{4}N_{2}$)).
Figure \ref{fig:re3} shows the permittivity and permeability of zolpidem
and caffeine (1H-imidazole ($C_{3}H_{4}N_{2}$)). All these permittivities
$\varepsilon_{r}$ are in the $xy$ plane, while permeabilities $\mu_{r}$
are in $z$ direction of the systems. We suppose that the life time
of the excited-state is $\tau=10$ns \cite{Tokuji09}. For pyrrole
($C_{4}H_{5}N$) the main contribution comes from the transition to
the fifth excited state, e.g. a electron transits from $\varepsilon_{3}$
to $\varepsilon_{5}$ with no flip. And the third transition makes
the major contribution to negative indexes of 1H-pyrazole ($C_{3}H_{4}N_{2}$).
While for zolpidem and caffeine, which hails from the transition to
the forth excited state. All these molecules display negative permittivity
and permeability simultaneously, thereby bring negative refraction
consequently. For negative magnetic response, the permeability of
AKB48 is much more obvious than that of the others (about $10$ times). While the result
of permittivity is reversed (about $3$ to $5$ times in $10^6$).  The phenomenon of negative
refraction of AM2201, JWH-018, JWH-073,
zolpidem and caffeine are of significant to AKB48. To achieve negative permittivity
and permeability simultaneously, the pickwidths of AKB48 in the figure
are wider than others. In other words,  the pickwidths reveals that AKB48 is much
easier to realize negative refraction at the present stage.

\section*{Conclusion and Discussion \label{sec:CONCLUSION}}

In this paper we put forward a new method to identify certain kinds
of psychoactive drugs which have the structures called split-ring resonators.
This configuration might induce negative permittivity and permeability
simultaneously, consequently negative refraction in the UV-vis region.
When light with certain frequency transmit through the transparent
media which has psychoactive drugs with negative indexes on it,
the refracted light goes on the opposite side comparing with the ordinary
case. Namely in common case, when the incident light transmit onto
the surface of media , the refracted light lies on the contrary side
of normal. But for negative refraction material, the refracted light
and incident light lie on the same side of normal. This phenomenon
made us capable of distinguishing certain families of psychoactive
drugs from others. The optical method has its advantage which is non-damaged
and would not cumber the detection of DNA in forensic science. Here
we use H\"{u}ckel model to deal with the systems. As a result, we
conclude that the main ingredients of synthetic cannabinoids, zolpidem
and caffeine can show various degree of negative refraction, thereby
can be distinguished from others. In psychoactive artificial cannabinoid
families e.g. AM-xxx, HU-xxx, JWH-xxx, CP xx, other small quantity
of ingredients e.g. AM-694, JWH-250 can also give the result of negative
refraction but they are unconspicuous. It is a remarkable fact that
the category of zepam which has a broken ring of heptagon could also
give negative refraction in all psychoactive drugs. But the category
of zepam is not a prevalent abused drug as synthetic cannabinoids
and others, therefore we only give analytic derivation and numerical
simulation of three simplified models.

\section*{Acknowledgement}

The research was supported by NSF of China under Grant No. 11575125. We thank for the useful discussion
of Pro.Wu-Sheng Dai from Department of Physics, Tianjin University.

\section*{Author contributions statement}

Yao Shen designed the project, wrote the main manuscript text  and did the calculations. Yu-Zhu Chen  reviewed the manuscript.

\section*{Additional information}

Competing financial interests: The authors declare no competing financial interests.

\end{document}